# LA POLÉMICA DEL MULTIVERSO

# THE MULTIVERSE CONTROVERSY

A. Gangui [a,b],*

a Universidad de Buenos Aires, Facultad de Ciencias Exactas y Naturales, Argentina.
b CONICET – Universidad de Buenos Aires, Instituto de Astronomía y Física del Espacio (IAFE), Argentina.

En este trabajo se ofrece una síntesis de las posiciones existentes en la literatura con respecto a las controversias sobre el multiverso. Luego de reseñar algunos elementos simples de la cosmología moderna y de sus limitaciones observacionales, presentaremos la historia de las ideas que llevó, primero, a la propuesta del principio antrópico y, años más tarde, a la posible existencia de dominios causalmente desconectados, eventualmente dotados de sus propias leyes y constantes fundamentales, no todos compatibles con la existencia de observadores, que fueron globalmente bautizados con el confuso nombre de universos paralelos o multiverso.

*Palabras clave: historia de la ciencia, astrofísica, epistemología.*

This paper offers a synthesis of the positions in the literature regarding controversies about the multiverse. After reviewing some simple elements of modern cosmology and its observational limitations, we will present the history of the ideas that led, first, to the proposal of the anthropic principle and, years later, to the possible existence of causally disconnected domains, eventually endowed with their own laws and fundamental constants, not all compatible with the existence of observers, which were globally baptized with the confusing name of parallel universes or multiverse.

## I. EL UNIVERSO OBSERVABLE Y SUS LÍMITES

Dado que aquí hablaremos sobre el universo (y también sobre lo que entendemos por multiverso, universos paralelos, etc.), trataremos de proponer una definición. Entendemos el "universo" como la totalidad, todo aquello que existe, existió o existirá algún día. De acuerdo con su etimología la palabra tiene un origen que se remonta seguramente a antes de la voz latina *universum* (todas las cosas, los seres, el mundo), que literalmente significa "convertido -o vuelto- en uno": de *unus* (uno) + *versus* (participio pasado de *vertere*: dar vuelta, convertirse, transformarse). Podemos entonces definir "universo" como "todo en uno, la totalidad". Más formalmente, en la cosmología relativista el universo es el espacio-tiempo único, conexo (es decir, conexo local y causalmente), poblado de campos y materia, del cual nuestro dominio cósmico (en expansión, observable) es tan sólo una parte.[1]

Aunque esta definición parece clara, los últimos años de la cosmología han logrado convulsionar la idea que teníamos del cosmos y a qué objeto o entidad se llama "universo". Hoy en día la mayoría de los cosmólogos acepta la teoría del Big Bang, según la cual hace unos 14 mil millones de años, el universo se hallaba en un estado primordial de altísima densidad y temperatura, que luego evolucionó -diluyéndose y enfriándose con la expansión- hasta el estado presente. Como es de imaginar, dado ese tiempo -inmenso pero no infinito- un observador no llega a ver más allá que la distancia que la luz ha podido recorrer desde aquel estado primordial, es decir unos 42 mil millones de años-luz, o sea unas tres veces el valor naif, dado que el universo se halla en expansión (aquí estamos considerando el modelo simple de Einstein-de Sitter del año 1932, que se expande como $R \sim t^{2/3}$, donde "R" es el factor de escala y "t" es el tiempo cósmico).[2]

Esta distancia máxima que logramos observar hoy es lo que comúnmente llamamos el "horizonte" de nuestro universo observable. Por supuesto, el universo no termina ahí. Es solo que, lo que se halla más allá de esa distancia no puede aún ser observado. Por razones de homogeneidad (el principio cosmológico), uno esperaría que más allá del horizonte existieran otros dominios en expansión, inobservables para nosotros, pero que también pertenecen a nuestro (gran) universo y que surgieron -al igual que nuestro dominio observable- como resultado del mismo "Big Bang". Como vemos, la presencia de dominios independientes (y causalmente separados) del nuestro es una predicción estándar de la cosmología moderna usual.

Hoy se considera que nuestro horizonte aumenta (algo más de) un año-luz por cada año que pasa. Por lo tanto, esos diferentes dominios causales, con el paso del tiempo podrían solaparse. Esto es así, si la expansión se va desacelerando. Sin embargo, si el universo se halla en aceleración -como las observaciones actuales indican- la situación cambia. Y esto no se reconoció hace poco. Ya en un trabajo de 1931, Eddington señaló

---

* *gangui@iafe.uba.ar*

que durante una expansión acelerada se llegaría a la situación en la cual[3]

*los objetos que se separan más rápidamente que la velocidad de la luz se aíslan de cualquier inferencia causal entre sí, de modo que al cabo de un tiempo el universo se convertirá virtualmente en una serie de universos desconectados que ya no tienen relación física entre sí.*

Recordemos que el modelo de Lemaître-Eddington espacialmente cerrado incluye una constante cosmológica positiva y se expande asintóticamente a partir del estado (estático) del modelo de Einstein. Vemos entonces que, dadas las condiciones adecuadas (una fase de aceleración), nuestro (gran) universo podría estar formado por múltiples regiones causales (más pequeñas, que Eddington llama "universos desconectados"), pero que no tienen -ni tendrán- relación causal con la nuestra.[4]

## II. DE LOS DOMINIOS CAUSALMENTE DESCONECTADOS A LOS MINI-UNIVERSOS

En su trabajo historiográfico, Kragh[5] reseña varias otras propuestas de universos formados por dominios independientes, sobre todo aquellas que se basan en las ecuaciones de campo de la Relatividad General. Por ejemplo, Pachner considera una colección de "universos" espacialmente cerrados sumergidos en un espacio de dimensión superior a las cuatro espaciotemporales usuales, de tal manera que sus hipersuperficies no se intersecan.[6] En ese sentido afirma que, aunque no hay interacción entre estos universos, eso no significa que no puedan coexistir.

Antes del advenimiento de la inflación cósmica, cuyas ideas ya estaban en el ambiente hacia fines de la década de 1970, pero tomaron vuelo con el trabajo de Guth,[7] las publicaciones sobre "muchos universos" no atraían mayormente la atención de los investigadores. Las versiones sobre la inflación "caótica" y "eterna", propuestas por Linde y Vilenkin poco tiempo más tarde, cambiaron este panorama.[8,9] En particular, en una presentación de 1982, en ocasión del workshop Nuffield, Linde sugería que luego de la inflación el universo quedaba dividido en una infinidad de sub-universos o burbujas separadas.

Ideas similares a esta fueron puestas en práctica por Gott, quien propuso un modelo en el cual se creaban, a partir de un universo inflacionario (con espacio de De Sitter), otros universos espacialmente abiertos (universos de Friedmann con curvatura k negativa), que resultaban causalmente disjuntos entre sí (y con el nuestro).[10] Más tarde, Sato y sus colaboradores propusieron una extensión de estas ideas, en la cual burbujas independientes quedarían atrapadas dentro del dominio en expansión inflacionaria. Con el tiempo, estas burbujas se evaporarían, al ser inestables y de alta energía, pero sólo después de haber formado universos hijos (Ref. 11, p. 539).

La idea de Linde, sin embargo, era más radical. En lugar de suponer la existencia de una singularidad inicial y un estado primigenio de altísima densidad y temperatura, propias de los modelos estándar del Big Bang, su modelo imaginaba una multitud de mini-universos, de los cuales ninguno sería el primero. Estos pequeños universos luego producirían otros mini-universos, llamados entonces universos-bebé, y así por siempre. Aunque no todos estos sub-universos habrían de sobrevivir, pues algunos podrían incluso colapsar debido a sus particulares parámetros característicos, el proceso de generación de universos seguiría por siempre (sería "eterno"). En palabras de Linde[12]

*el universo es una entidad que existe eternamente y que se auto reproduce, que está dividida en muchos mini-universos mucho mayores que nuestra porción observable, y que las leyes de la física de las bajas energías e incluso la dimensionalidad del espacio-tiempo pueden ser diferentes en cada uno de estos mini-universos.*

Un corolario que se deduce de esto es que no existiría un único inicio para el universo como un todo, y tampoco habría de haber un final.

Como consecuencia de los modelos inflacionarios, entonces, nuestro dominio observable sería tan solo una pequeña parte de una burbuja cósmica homogénea que luego sufrió una fase de expansión exponencial extra-rápida, impulsada por un campo escalar (el "inflatón"). Por supuesto, esto explica por qué nuestro universo observable, desde épocas tempranas, es tan suave y homogéneo, con un valor de la curvatura espacial tan próximo a cero, característico de la densidad crítica que separa universos en expansión eterna de aquellos que eventualmente recolapsan (Ref. 2, p. 226). Algunos modelos, además, predicen -como vimos- que existen muchas otras burbujas con muy diferentes propiedades. En la variante "eterna" de la inflación, el universo continuamente se auto-reproduce, lo que lleva a una infinidad de sub-universos que se extienden en el espacio y en el tiempo.[13,14]

## III. IDEAS ANTRÓPICAS

Las diferentes propiedades para los diferentes sub-universos generados durante la inflación eterna son quizá reminiscentes de las anteriores y muy influyentes ideas de Carter,[15] cuando propuso su formulación original del principio antrópico ("débil") según la cual, por ejemplo, la edad del universo no puede ser arbitraria, sino que debe ser compatible con el requerimiento de que la vida tenga tiempo suficiente para desarrollarse.

Aunque Carter no fue el primero en mencionar estos temas, en su trabajo se observaba que las características básicas de las galaxias, las estrellas, los planetas y el mundo microscópico subyacente están determinadas esencialmente por unas cuantas constantes microfísicas y por los efectos de la gravitación. Varios aspectos de nuestro universo -algunos de los cuales parecen ser los pre-requisitos para la evolución de cualquier forma de vida- dependen de una manera muy delicada de ciertas "coincidencias" evidentes entre las constantes físicas.[16]

La hipótesis antrópica, sin embargo, solo tenía poder explicativo si venía asociada con la existencia de un *ensemble* de universos, con diferentes combinaciones de constantes fundamentales y parámetros cosmológicos (o mejor, de todas las combinaciones concebibles). El nuestro sería entonces tan solo una realización posible dentro de dicho grupo (eventualmente infinito) de universos: un universo en el que las condiciones físicas -las leyes y las constantes- serían las adecuadas para el desarrollo de observadores (y de la conciencia). En palabras de Carter (Ref. 17, p. 348):

*El principio antrópico es un punto medio entre el antropocentrismo primitivo de la era pre-copernicana y la antítesis igualmente injustificable de que ningún lugar o tiempo en el Universo puede ser privilegiado en modo alguno.*

Con un poco más de detalle y en dos de sus versiones más relevantes: el principio antrópico "débil" acepta las leyes de la naturaleza y las constantes físicas como dadas, y afirma que la existencia de observadores impone entonces un efecto de selección sobre dónde y cuándo observamos el universo. Por su parte, el principio antrópico "fuerte", al menos en el sentido que le dieron Carr y Rees,[16] sugiere que la existencia de observadores impone restricciones sobre las propias constantes físicas, vale decir, estas constantes no pueden tomar valores arbitrarios.[17]

Del principio antrópico se deduce entonces que las "coincidencias" en los valores que se miden de los parámetros cosmológicos y de las constantes fundamentales de la física no son tales, sino que el universo observable está condicionado por la existencia de la vida compleja (y de observadores humanos).

Se pueden dar varios ejemplos, útiles para comprender la importancia de estas ideas en cosmología. Uno se basa en el parámetro "Q", que se define como la amplitud de las fluctuaciones cosmológicas de densidad de materia. De acuerdo al análisis desarrollado por Tegmark y Rees,[18] el valor de este parámetro no puede ser muy diferente del que fue hallado (del orden de $Q \sim 10^{-5}$, detectado por el satélite COBE), o de lo contrario el universo actual habría resultado muy distinto del que observamos. Otro caso notorio de la coincidencia mencionada es la relación que se da entre algunos números adimensionales relevantes para la física fundamental, en particular los que surgen de la interacción de un electrón y un protón (el conocido enfoque de Dirac de 1937). Estos ejemplos ya fueron discutidos con cierto detalle en Ref. 19.

Vemos entonces que -en general- el principio antrópico es esencialmente un "principio de selección", que señala que el universo observable está condicionado por la existencia de formas de vida compleja. Como principio de selección exige que las condiciones iniciales, los parámetros y las constantes fundamentales estén "ajustados" adecuadamente como para permitir la existencia de observadores (conscientes).

La combinación de estas ideas y avances teóricos (es decir, la generación de sub-universos en la inflación eterna, que describimos antes, y la propuesta antrópica de un *ensemble* de realizaciones con todas las posibles variaciones de parámetros y constantes) tardó un tiempo en convertirse en una línea de investigación genuina. Esta situación, sin embargo, fue cambiando en los últimos años.

## IV. CLASIFICACIÓN DE MULTIVERSOS: EJEMPLOS DE LA HISTORIA DE LA CIENCIA

Antes de desarrollar los modelos de "muchos universos" más recientes que han ido surgiendo en la literatura, conviene detenernos e intentar hacer una clasificación. De acuerdo a Gale los modelos pueden separarse en tres tipos cualitativamente distintos: universos temporalmente múltiples, universos espacialmente múltiples, y finalmente universos múltiples en otras dimensiones.[11]

Los primeros (que también podríamos denominar: multiversos temporales) son quizá los más antiguos como motivo de especulación, pues (aunque pueda sonar anacrónico) su historia podría remontarse a los presocráticos. Anaximandro, por ejemplo, ya en el siglo VI a.C., propuso ideas que sugieren un cosmos cíclico en el tiempo, con la destrucción del universo "viejo" al finalizar cada período, y la creación de uno nuevo al comenzar el siguiente. Al menos así nos lo relatan los comentaristas antiguos de su pensamiento en lo que quizá sería una serie infinita de ciclos, generados por la acción de los "opuestos primarios", llamados metafóricamente "amor" y "lucha" o con nombres similares. Por ejemplo, Pseudo Plutarco dice que, según Anaximandro (Ref. 20, p. 114)

*lo infinito es la causa de la generación y destrucción de todo, a partir de lo cual se segregan los cielos y en general todos los mundos, que son infinitos. Declara que su destrucción y, mucho antes, su nacimiento se producen por el movimiento cíclico de su eternidad infinita.*

Esto, por supuesto, nos recuerda algunos modelos de la cosmología más reciente, que proponían que a la actual fase de expansión le seguiría una fase de contracción (y eventualmente un "Big Crunch"), en lo que sería una serie análoga de ciclos de expansión/contracción que podría no haber tenido inicio, ni tendría un final. Como es claro de esta descripción, estos "universos" jamás co-existían en el tiempo: la completa destrucción de uno era seguida por la creación de uno nuevo (por eso son multiversos temporales). Una reencarnación moderna de estas ideas la encontramos en los trabajos de los universos cíclicos de Steinhardt y Turok.[21,22]

Muy diferente es lo que sucede con los "universos espacialmente múltiples" (o multiversos espaciales), pues en este caso los diferentes "universos" existen simultáneamente (aunque aquí quizá deberíamos llamarlos "mundos" y no universos, de acuerdo a la definición que dimos más arriba, donde el universo lo contiene todo). Muy probablemente, uno de los primeros ejemplos que nos viene a la memoria es el de los diálogos *Sobre el infinito universo y los mundos* de Giordano Bruno, quien en el siglo XVI propuso una infinidad de mundos, compuestos por sus soles y sistemas planetarios habitados, distribuidos por un espacio infinito. (Recordemos que Bruno distinguía claramente entre universo, mundos y Tierras, con la posibilidad de la existencia de infinitos mundos.)

Por último, los "universos múltiples en otras dimensiones" comenzaron con ideas de tipo eminentemente religioso en la pluma del filósofo y teólogo Leibniz. En sus escritos se invocaba la capacidad de Dios de crear no uno sino muchos (infinitos) mundos, y no hacerlo en el espacio o a lo largo del tiempo, sino "en paralelo", donde una posible realización se distinguía de la otra por detalles infinitesimalmente pequeños. No iremos a los detalles, pero ideas de este tenor volvieron a surgir, esta vez en el ámbito científico de mediados del siglo XX, luego del desarrollo de la mecánica cuántica, con la interpretación de "muchos mundos" de Everett.[23] Recordemos que esta aproximación buscaba evitar tener que invocar el colapso de la función de onda, un paso esencial en el proceso de medición según la interpretación de Copenhague. En su lugar, la interpretación de Everett supone que el universo "se divide" cada vez que se efectúa una medición, de tal manera que rápidamente se va generando una cantidad inmensa de "universos paralelos".

## V. ¿QUÉ ES UN UNIVERSO? (COMO MIEMBRO DEL MULTIVERSO)

Aunque al inicio ya presentamos una posible definición, en lo discutido hasta aquí no nos detuvimos a definir precisamente qué *características* debe tener un "universo" (a veces también confundido con mundo o dominio causalmente desconectado), pero ahora como parte de un multiverso. Tengamos en cuenta que, como lo venimos describiendo, una serie, o un grupo, o un ensemble de universos es lo que, en fin de cuentas, constituye nuestro multiverso.

Gale (Ref. 11, p. 534) propone que los universos miembros de un multiverso deben satisfacer las dos condiciones siguientes:

1. *completitud*: los universos deben ser completos, incluyendo la posibilidad de albergar formas de vida inteligente;

2. *separación*: los diferentes universos deben ser independientes y separados.

Un "multiverso temporal" caracterizado, por ejemplo, por una serie cíclica de creaciones y destrucciones, como es el caso del universo de Anaximandro (si el anacronismo es tolerable) verificaría la primera condición (completitud), ya que cada nuevo universo creado, supuestamente, poseía todas las cualidades posibles. Sin embargo, cada fase no se hallaba completamente separada de las demás -no era independiente-, pues las mismas cualidades de un universo se repetían en todos ellos. Podemos también pensar en otros ejemplos en donde se reitera esta situación, vale decir donde los universos individuales son completos pero donde la separación entre ellos no es total. Para satisfacer el requerimiento de "separación", los universos deben estar causalmente separados/desconectados, ser independientes, y no debe ser posible el intercambio de señales o información. Dentro de la cosmología contemporánea, en algunos modelos del estilo de Steinhardt y Turok se propone que los ciclos presente y pasados están físicamente desconectados (o sea, que no hay información que pase entre ellos), aunque en otros sí hay un cierta memoria del ciclo anterior.[21]

Por otra parte, volviendo al siglo XVI, en los diálogos de Giordano Bruno uno de sus personajes, Filoteo, menciona una frase que hoy ya se ha vuelto célebre (Ref. 24, p. 111):

*Allí innumerables estrellas, astros, globos, soles y tierras se perciben con los sentidos, y otros infinitos se infieren con la razón. El universo inmenso e infinito es el compuesto que resulta de tal espacio y de tantos cuerpos en éste comprendidos.*

De este extracto del inicio del diálogo tercero, podemos inferir que aquellos mundos que "se perciben con los sentidos" no satisfacen la condición de separación (con respecto a nuestro mundo, por ejemplo). Pero aquellos infinitos mundos que "se infieren con la razón", muy probablemente sí la verifiquen. Quizá podríamos sugerir que el Nolano concebía una idea de muchos mundos (o multiverso) espacial.

En resumen, ¿qué es un universo, como miembro de un multiverso? Podemos afirmar que es una entidad, un mundo o sistema de mundos que es completo y que está causalmente separado de todos los demás. Veremos que, en la práctica, es muy difícil, si no imposible, trabajar con esta definición, y de hecho son pocos los cosmólogos que lo hacen. En cualquier caso, se precisa de un mecanismo o proceso "meta-cósmico" para lograr fabricarlos,[1] como es el caso en el marco de la inflación caótica o eterna.

## VI. CLASIFICACIÓN DE MULTIVERSOS EN LA COSMOLOGÍA RECIENTE

Podemos ahora preguntarnos ¿qué es un multiverso? Vale decir, ¿cómo se clasifican los posibles multiversos de acuerdo a la cosmología actual? En base a lo ya desarrollado, un multiverso es un conjunto de universos pero con ciertas propiedades.

Según Tegmark,[25] los multiversos pueden dividirse en cuatro tipos o clases diferentes:

1. Un universo con expansión acelerada que termina generando múltiples regiones causalmente desconectadas, pero donde todos estos dominios surgieron de un mismo "inicio" (en las épocas embrionarias de un Big Bang), es lo que se califica como multiverso de clase-I. Esta clase, como vimos, no es nada controversial.

2. La clase-II, en cambio, es aquella en donde hay varios big bangs, como sucede genéricamente en los modelos de la inflación eterna de Linde, o bien en los modelos oscilantes de Steinhardt y Turok, y donde los distintos universos comparten las mismas leyes físicas, pero son regidos por diferentes constantes fundamentales. (Aquí se combinan los modelos de multiversos temporales y espaciales de Gale que ya hemos mencionado.)

3. La clase-III, por su parte, está reservada a aquellos multiversos que surgen en el marco de la interpretación de muchos mundos de Everett de la mecánica cuántica.

4. Finalmente, la jerarquía termina con los multiversos de clase-IV, que agrupa universos que no sólo no comparten leyes o constantes, sino que, además, podrían tener estructuras matemáticas diferentes. Es decir, aquí se postula que todo universo matemáticamente posible, tiene realidad física y es uno de los integrantes de ese multiverso. A esta propuesta, que de más está decir no cuenta con muchos adeptos, a veces se la llama "democracia matemática".

Pero los universos de la inflación eterna -supuestamente- tienen todos un origen causal común y, además, comparten el mismo espacio-tiempo y sus leyes físicas (decimos "supuestamente" porque Linde afirmaba que en estos modelos *no* había un origen). Por esta razón, para algunos investigadores no forman un multiverso con integrantes completamente desconectados entre sí (la condición de "separación" en la clasificación de Gale, pero ahora más restrictiva, cf. Ref. 1). Lo mismo puede afirmarse para los modelos cíclicos de Steinhardt y Turok. Estos dos son los multiversos (pertenecientes a la clase-II) que más se han trabajado en los últimos años, y que con mayor profundidad fueron estudiados para extraer predicciones sobre las constantes fundamentales, y especialmente sobre la constante cosmológica. Según Ellis y colaboradores,[1] al tener una cierta conexión (como mencionamos antes: origen causal común y espacio-tiempo común), se ha propuesto cambiarles el nombre y llamarlos "universo multi-dominio", dejando el nombre "multiverso" para aquellos conjuntos de universos estrictamente disjuntos y causalmente desconectados (y que ni siquiera compartan nuestro mismo espacio-tiempo). A pesar de ser una propuesta bastante coherente, quizá como era de esperarse, muy pocos son capaces de resistir a la tentación de emplear la palabra "multiverso" en el título de sus *papers*.

Como menciona Kragh,[5] en el pasado los físicos consideraban que las leyes de la naturaleza eran únicas y constituían los principios fundamentales a partir de los cuales se podían modelar los fenómenos naturales. Hoy, en cambio, los cosmólogos adeptos al multiverso consideran que no hay nada especial en las leyes que rigen el universo; estas serían meras "leyes derivadas", locales y permitidas antrópicamente, de manera tal de ser compatibles con la existencia de la vida como la conocemos. Estas leyes, entonces, -ya conocidas o aún por descubrir- dejan de ser necesarias y pasan a ser contingentes, al igual que muchos de los valores de los parámetros físicos y cosmológicos que determinan nuestro universo.

### VII. ESTATUS CIENTÍFICO Y LA POSIBILIDAD DE HACER PREDICCIONES

Llegados a este punto se presenta una cuestión de no poca importancia, pues: ¿basta con verificar que estos -quizás innumerables- universos posibles sean consistentes con las leyes básicas de la física-matemática? ¿O es necesario que sea posible detectarlos (observarlos) en la forma usual, empírica? La pregunta que ahora nos hacemos es ciertamente fundamental: la cosmología del multiverso, ¿es realmente una ciencia?

George Ellis, uno de los más fervientes detractores, sugiere que la afirmación de que existe un multiverso -compuesto de universos causalmente disjuntos- es metafísica y que, por su propia naturaleza, jamás podrá convertirse en una afirmación científica. Podríamos también reformular las preguntas anteriores diciendo: ¿deberemos por siempre conformarnos con satisfacer criterios "no-fácticos" (por ejemplo, la coherencia interna de una teoría) y relegar en cambio los criterios "fácticos" (vale decir, la contrastabilidad con la experiencia y la observación) cuando abordamos estos modelos de muchos universos?

Aunque sabemos que la mayoría de los filósofos hoy se aparta del empirismo tradicional, podemos preguntarnos sobre qué manera de falsar la teoría -en el sentido popperiano- tenemos a nuestra disposición. Siguiendo a Kuhn,[26] recordemos los cinco criterios estándar para evaluar cuan satisfactoria es una teoría científica: precisión, consistencia interna y externa (coherencia), amplitud de alcance, simplicidad, y fecundidad o productividad.

La primera condición (la precisión o exactitud) se relaciona con el poder empírico de una teoría, esto es, con el hecho de que existan consecuencias deducibles de la teoría que deben estar en acuerdo con el resultado de los experimentos y las observaciones. Aunque existen sutilezas en la implementación de estos criterios para la cosmología moderna,[27,28,29] hay un consenso general que selecciona al primero de ellos (la precisión) como uno fundamental, cuya verificación (o refutación) frente a la experiencia otorga estatus científico a una teoría. La confrontación contra el experimento o la observación (la falsación popperiana) es una condición necesaria, aunque no suficiente, como criterio de demarcación entre teorías científicas y aquellas que no lo son. (Aunque Popper[30] dejó en claro que no le asignaba ningún valor absoluto al criterio de falsación y no lo consideraba una definición de ciencia.)

Por supuesto, no todos los defensores del multiverso se ponen de acuerdo en qué representa en verdad la contrastabilidad (o incluso la precisión) de una teoría o cuan importante es este criterio en comparación con los demás. Por ejemplo, la simplicidad y la consistencia interna de algunos modelos de multiverso, ¿podrían ser más importantes y ser priorizadas frente a la verificación empírica? Susskind sugiere que la misma verificación no-empírica que avala la teoría de cuerdas (coherencia y consistencia matemática) puede aplicarse también al multiverso.[31]

### VIII. MUCHOS UNIVERSOS, MUCHA POLÉMICA

Una crítica usual que se le hace a las ideas antrópicas es que el célebre "principio" no hace predicciones comprobables. Por lo tanto no es falsable, y en consecuencia no es parte de la ciencia.

Sin embargo, Vilenkin afirma que los modelos antrópicos pueden ser sometidos a tests observacionales, y pueden ser ratificados o falsados con un adecuado nivel de confianza.[32] Este autor sugiere que las constantes de la física pueden, en realidad, ser variables estocásticas, y tomar diferentes valores en diferentes partes del megauniverso. Apoya sus conclusiones en las estimaciones que se han venido haciendo, precisamente, sobre el valor de la constante cosmológica y en su distribución sobre el conjunto de universos. También señala que las predicciones son de naturaleza eminentemente estadística (en función de distribuciones de probabilidad) y poseen una varianza intrínseca que no puede reducirse, a diferencia de lo que sucede con las observaciones astronómicas y los desarrollos teóricos, que pueden ser refinados con el paso del tiempo. Señalemos, sin embargo, que su análisis se aplica a un ensemble de dominios separados que no satisface las condiciones arriba expuestas para un multiverso.

Por su parte, Ellis argumenta que el concepto de multiverso no es único ni está bien definido.[27] Pueden existir muchas y muy diferentes realizaciones y, como corolario, aún no tenemos una adecuada teoría sobre el multiverso. También afirma que decir que "todo lo que puede suceder, en efecto sucede", como apuntan muchos autores, no especifica un multiverso único. A pesar de lo coherente de su postura, el estilo más conservador de Ellis no es seguido por todos. En contraposición, Vilenkin afirma que sí existe una teoría del multiverso y que esta es justamente la inflación eterna.

Nuestra opinión es que, como acabamos de señalar, la mayoría de los cálculos de probabilidad que se realizan sobre un ensemble de universos, en realidad no se aplican al multiverso como conjunto de universos estrictamente disjuntos y causalmente desconectados (la definición de Ellis), sino a un conjunto de dominios suficientemente separados, pero que comparten un mismo origen y espacio-tiempo.

Y esta última idea sí es verificable. Hay ejemplos, en efecto, en los que la posible existencia de un modelo con múltiples dominios puede ser falsada. Es el caso cuando se propone que el nuestro es un "universo pequeño", espacialmente compacto, donde un observador genérico ya ha visto todo el camino a su alrededor, es decir, donde un rayo de luz ha tenido tiempo suficiente de dar la vuelta al universo. En este caso, el universo se cierra sobre sí mismo en un solo dominio, del estilo Friedmann-Robertson-Walker, por lo que no puede haber otros dominios que estén causalmente desconectados de nosotros. Esta sería una manera de refutar la idea de múltiples dominios de la inflación caótica. Sin embargo, nada podemos decir sobre el multiverso, si en verdad existe, pues este último no puede ser verificado o falsado observacionalmente.[33]

## IX. CONCLUSIONES

De acuerdo a Page,[34]

*uno no puede poner a prueba científicamente una teoría que realiza predicciones sobre lo que es inobservable. Pero uno sí puede testear una teoría que hace uso de entidades no-observables para explicar y predecir aquellas que sí podemos observar.*

Esta idea, aplicada al contexto que venimos discutiendo, sin duda, propone la aceptación de los muchos universos (inaccesibles) como entidades teóricas que nos permiten predecir los valores de, por ejemplo, las constantes físicas que sí podemos medir. Otro interrogante que excede a la ciencia observacional es el origen de las verdaderas leyes de la física. Pues, si en efecto es verdad -como lo proponen algunos modelos- que estas leyes varían en los diferentes universos a lo largo y ancho del multiverso, deberían existir "meta-leyes" (superiores) que definieran de qué manera se determinan las leyes derivadas (más locales) de cada lugar (a menos que en el nivel meta rija el puro azar). En esta situación, además, podemos preguntarnos acerca del rol del observador. O más precisamente, qué significa ser un observador "típico" (como acostumbramos considerar en la ciencia tradicional) en un sistema en el cual todo posible observador va a aparecer, y encima un número infinito de veces.

Para los escépticos, sin embargo, la idea que queda de todo esto es que, al no haber conexión causal entre los distintos universos propuestos y nuestros aparatos de observación, la supuesta existencia de un verdadero multiverso es, cuando más, una suposición metafísica. Como señalamos, no puede ser parte de la ciencia "fáctica" usual, ya que esta exige pruebas experimentales u observacionales para descartar los modelos erróneos (la contrastabilidad con la experiencia y la observación). Muchos autores, sin embargo, no encuentran nada de malo en que existan argumentos metafísicos, ya que ese es un abordaje genuino del problema, pero este no debe confundirse con la metodología de las ciencias experimentales.

Para terminar, recuperamos uno de los argumentos de Ellis, quien se cuestiona si realmente existe un multiverso. Y su respuesta es que, simplemente, no sabemos, y probablemente nunca lo sepamos (Ref. 33, p. 406). Al final,

*la creencia en un multiverso será siempre una cuestión de fe de que los argumentos lógicos propuestos dan la respuesta correcta en una situación donde una comprobación observacional directa es inalcanzable y donde la física subyacente supuesta no es testeable.*

## Agradecimientos

## X. REFERENCIAS